**Changes in the *b* value in and around the focal areas of the *M*6.9 and *M*6.8 earthquakes off the coast of Miyagi prefecture, Japan, in 2021**

K. Z. Nanjo[1,2,3,*], A. Yoshida[2]

1. Global Center for Asian and Regional Research, University of Shizuoka, 3-6-1, Takajo, Aoi-ku, Shizuoka, 420-0839, Japan

2. Center for Integrated Research and Education of Natural Hazards, Shizuoka University, 836, Ohya, Suruga-ku, Shizuoka, 422-8529, Japan

3. Institute of Statistical Mathematics, 10-3 Midori-cho, Tachikawa, Tokyo, 190-8562, Japan

*Corresponding author: K. Z. Nanjo

Email: nanjo@u-shizuoka-ken.ac.jp; Tel: +81-54-5600; Fax: +81-54-5603

ORCID ID: https://orcid.org/0000-0003-2867-9185

# Abstract

We investigated changes in the $b$ value of the Gutenberg-Richter's law in and around the focal areas of earthquakes on March 20 and on May 1, 2021, with magnitude ($M$) 6.9 and 6.8, respectively, which occurred off the Pacific coast of Miyagi prefecture, northeastern Japan. We showed that the $b$ value in these focal areas had been noticeably small, especially within a few years before the occurrence of the $M$6.9 earthquake in its vicinity, indicating that differential stress had been high in the focal areas. The coseismic slip of the 2011 Tohoku earthquake seems to have stopped just short of the east side of the focus of the $M$6.9 earthquake. Furthermore, the afterslip of the 2011 Tohoku earthquake was relatively small in the focal areas of the $M$6.9 and $M$6.8 earthquakes, compared to the surrounding regions. In addition, the focus of the $M$6.9 earthquake was situated close to the border point where the interplate slip in the period from 2012 through 2021 has been considerably larger on the northern side than on the southern side. The high-stress state inferred by the $b$-value analysis is concordant with those characteristics of interplate slip events. We found that the $M$6.8 earthquake on May 1 occurred near an area where the $b$ value remained small, even after the $M$6.9 quake. The ruptured areas by the two earthquakes now seem to almost coincide with the small-*b*-value region that had existed before their occurrence. The *b* value on the east side of the focal areas of the *M*6.9 and *M*6.8 earthquakes which corresponds to the eastern part of the source region of the 1978 off-Miyagi prefecture earthquake was consistently large, while the seismicity enhanced by the two earthquakes also shows a large *b* value, implying that stress in the region has not been very high.

## Keywords

Statistical analysis; Earthquake dynamics; Seismicity and tectonics; Earthquake interaction, forecasting, and prediction; Computational seismology; Stresses; Spatial analysis; Time series analysis; Subduction zone processes

## Main Text

**Introduction**

An earthquake of magnitude (*M*) 6.9 occurred on March 20, 2021, at 18:09, on the Pacific coast of Miyagi prefecture, northeastern Japan, and successively, an *M*6.8 earthquake occurred on May 1, 2021, at 10:27, at about 50 km south of the *M*6.9 earthquake (Figures 1a,b) (Earthquake Research Committee (ERC), 2021a,b,c). Both earthquakes were located on the periphery of the slip region of the 2011 *M*9.0 Tohoku earthquake (Figure 1a), almost corresponding to the down-dip end of the interplate coupling zone between the overriding continental and the subducting Pacific plates (Igarashi et al., 2001). As can be seen in Figure 1a, the foci of the two 2021 earthquakes were located in the area where the coseismic slip (orange contours) of the 2011 Tohoku earthquake was relatively large and the afterslip (black contours) during 7 months just after the earthquake occurrence was small compared to the surrounding region (Ozawa et al., 2012). Interestingly, it was reported that the focus of the *M*6.9 earthquake was situated at the border point where the interplate slip during the period from 2012 through 2021 inferred by the analysis of repeating earthquakes was large on the northern side and notably small on the southern side (Tohoku University, 2021b).

When the *M*6.9 earthquake occurred on March 20, the ERC (2021a,b) pointed out that the focal area was located in the western part of the source region of the so-called off-Miyagi prefecture earthquake (Figure 1b), an interplate earthquake that has occurred sequentially at intervals of about 38 years, the most recent one being the *M*7.4 earthquake in 1978 (green contours in Figures 1a,b; Yamanaka and Kikuchi, 2004). Note that the focal area of the *M*6.8 earthquake on May 1 was located at the west of the slip region of the *M*7.2 earthquake that occurred at off-Miyagi prefecture in 2005 (purple contours in Figures 1a,b; Yaginuma et al., 2006). There was an overlap with the southeastern part of the source region of the 1978 off-Miyagi prefecture earthquake (Figure 1b).

Here, we report the results of our analysis on spatio-temporal changes in the *b* value of the Gutenberg-Richter's (GR) law (Gutenberg and Richter, 1944) in and around the focal areas of the *M*6.9 and *M*6.8 earthquakes and in the source regions of the 1978 and 2005 off-Miyagi prefecture earthquakes, and discuss the implications of these results, noting the stress state on the plate interface in these regions.

**Method**

We exploited the GR law, $\log_{10}N = a - bM$, where $N$ is the number of events equal to or above *M*, and *a* and *b* are constants (Gutenberg and Richter, 1944). Globally, on average, $b \sim 1$, but locally, $b$ values show substantial spatial and temporal variation. In some cases, the proportion of earthquakes with large magnitudes is higher ($b < 1$), in others, the proportion of earthquakes with small magnitudes exceeds the average expectation ($b > 1$) (Figures 2-5).

To estimate $b$ values consistently over space and time, we employed the EMR (Entire-Magnitude-Range) technique (Woessner and Wiemer, 2005), which simultaneously calculates the completeness magnitude $M_c$, above which all events are considered to be detected by the referential seismic network. EMR applies the maximum likelihood method represented by Eq. (6) of Utsu (1999) (e.g., Aki, 1965; Utsu, 1965) when computing the $b$ value to events with magnitudes greater than $M_c$. We calculated the $b$ values (Figures 2-5), provided that we found a minimum of 20 events with magnitudes greater than $M_c$ for a given sample. We evaluated the uncertainty of the maximum-likelihood estimates of the $b$ value, as described in Shi and Bolt (1982). The difference in $b$ is not considered to be significant if the test proposed by Utsu (1992, 1999) is not passed. If $\log P_b$, the logarithm of the probability that the $b$ values are not different, is equal to or smaller than -1.3 ($\log P_b \leq -1.3$), then the difference in $b$ is significant (Schorlemmer et al. 2004; Nanjo and Yoshida, 2017). A fit of the GR law to observations for three circle areas is given in Figure 3c, where the $b$ value was estimated by the maximum likelihood method (Aki, 1965; Utsu, 1965, 1999), rather than a coefficient of the GR law. The $b$ value is smaller for the circle 1 ($b = 0.4 \pm 0.1$) than for the circle 2 ($b = 1.6 \pm 0.1$), taking an intermediate value for the circle 3 ($b = 0.9 \pm 0.2$), indicating the significant difference in $b$ among the three circle areas. This significance is further supported by the Utsu test (Utsu, 1992, 1999; Schorlemmer et al. 2004; Nanjo and Yoshida, 2017), revealing log$P_{b1,3}$ (the logarithm of the probability that the $b$ values for the circles 1 and 3 are not different) to be -3.6, and log$P_{b2,3}$ for the circles 2 and 3 to be -4.5 (Figure 3).

The $b$ value is known to be sensitive to differential stress and its inverse correlation with differential stress has been evidenced by many laboratory and field studies (Mogi, 1963; Scholz, 1968, 2015; Lei, 2003; Schorlemmer et al., 2005; Goebel et al., 2013). Investigation into space-temporal variation in $b$ values to probe the stress state in the Earth's crust (Smith, 1981; Schorlemmer et al., 2005; Narteau et al., 2009) has been applied to locate asperities (Hirose et al., 2002; Yabe, 2003; Tormann et al., 2015; Nanjo and Yoshida, 2018), and to estimate frictional properties (e.g., Sobiesiak et al., 2007; Ghosh et al., 2008) on the plate interface along subduction zones. Foreshocks have been known to show small *b* values (Suyehiro et al., 1964; Gulia and Wiemer, 2019). Patches with small *b* values on active faults have been observed to coincide with locations of subsequent large earthquakes (Schorlemmer and Wiemer, 2005; Nanjo et al., 2016, 2019; Nanjo, 2020).

**Data**

We used the Japan Meteorological Agency (JMA) earthquake catalog, which includes earthquakes since 1919 in and around Japan. Our analysis was based on earthquakes with $M \geq 0.5$ during the period from January 1, 2012 through May 15, 2021, in a depth range of 0-100 km in the study region (Figure 1b). We did not consider the period from immediately after the $M$9.0 Tohoku earthquake on March 11, 2011 to the end of 2011, avoiding the time period with large temporal variation in *b*. Most pre-shocks and aftershocks of the two $M$6.9 and $M$6.8 earthquakes in 2021 in the study region (Figure 1b) occurred on and around the plate interface between the continental plate and the subducting Pacific plate (Figure 3), so our primary approach was to develop maps about the $b$ value and its cross-sections.

A $b$-value analysis is critically dependent on a robust estimate of completeness of the processed earthquake data. In particular, underestimates in $M_c$ lead to systematic underestimates in $b$ values. We always paid attention to $M_c$ when assessing $M_c$ at each node and each time window (Figures S1-S3 of Additional file 1). As discussed in other studies using the JMA catalog (Nanjo et al., 2010; Schorlemmer et al., 2018), $M_c$ in offshore regions is expected to increase with distance from the coast, and $M_c$ that had once increased during the period of an early aftershock sequence of the Tohoku earthquake, decreased afterward (Figure S1 of Additional file 1). One of the reasons for this change in $M_c$ was due to a change in the criterion for creating the JMA earthquake catalog in order to avoid the loss of integrity of work to determine hypocenters and the magnitudes of earthquakes (JMA, 2012; Schorlemmer et al., 2018).

**Results**

Figure 2a represents $b$ maps in three periods: from January 1, 2012 through December 31, 2014, from January 1, 2015 through December 31, 2017, and from January 1, 2018 through March 20, 2021 before the occurrence of the $M$6.9 earthquake. The left panels of Figures 3a,b are east-west cross-sectional views of $b$ values, crossing the $M$6.9 and $M$6.8 hypocenters, respectively, based on seismicity (black circles in the right panels) from January 1, 2018 to immediately before the $M$6.9 earthquake (March 20, 2021, at 18:08). It is notable (Figures 2 and 3) that the hypocenter of the $M$6.9 earthquake is located within an especially small-$b$-value structure (blue to purple) along the plate interface. The $b$ value around the hypocenter of the $M$6.8 earthquake is also very small.

Note that the decrease in the $b$ value in the area near the foci of the two 2021 earthquakes became conspicuous in later periods before the $M$6.9 earthquake (Figure 2). In the first period, the local $b$ value for the seismicity around the eventual $M$6.9 hypocenter was $b = 0.8 \pm 0.3$ (light blue), nearly a global average ($b \sim 1$) (Figure 3d). However, the $b$ value in the second period became $0.6 \pm 0.2$ (blue), and it decreased to a low value ($0.4 \pm 0.1$) in the last period. As shown in Figure 2b, the decrease in the $b$ value around the focus of the 2021 $M$6.9 earthquakes in the third period relative to the first period was statistically significant (Utsu, 1992, 1999; Schorlemmer et al., 2004; Nanjo and Yoshida, 2017). The decrease in the $b$ value in region A, a region surrounding focal areas of the 2021 $M$6.9 and $M$6.8 earthquakes, is also seen from Figure S4 of Additional file 1, although the decrease in $b$ is not clear in the focal area of the $M$6.8 earthquake as that in the focal area of the $M$6.9 earthquake (Figures S5 and S6 of Additional file 1).

Incidentally, it might be worth noting that the $b$ values in the middle panel of Figure 2a are small in the focal area of the future $M$6.2 earthquake that occurred in 2020 (in the bottom panel of Figure 2a).

Figure 4a shows a $b$ map calculated for earthquakes that occurred during the period between the $M$6.9 and $M$6.8 earthquakes. It is notable that the $b$ values in most parts of the focal regions of the two earthquakes, especially the southern part of the focal region of the $M$6.9 quake and the northern part of the focal region of the $M$6.8 quake in that period, were as small as those before the occurrence of the $M$6.9 earthquake, as is demonstrated by Figures 4b,c. It might also be noteworthy that the $b$ value for the seismicity induced in the eastern part of the source area of the 1978 off-Miyagi prefecture earthquake (green contours) seems to have increased after the $M$6.9 earthquake.

We also examined the temporal change in the $b$ value in regions A and B shown in Figure 5a during the period from January 1, 2018 through May 15, 2021. Similar to region A, region B corresponds to the eastern part of the source region of the 1978 off-Miyagi prefecture earthquake with $M$7.4, which includes the source area of the 2005 $M$7.2 earthquake that occurred at off-Miyagi prefecture. It is clear (Figures 5b,c) that the $b$ value in region A had been consistently small before the occurrence of the two earthquakes in 2021, and it is to be noted (Figure 5c) that the small $b$-value state in region A continued after the occurrence of the $M$6.8 earthquake on May 1. This seems to indicate that a highly-stressed state continued after the $M$6.8 earthquake in the area where the $b$ value had been small before the $M$6.9 earthquake. Some possible interpretations of this result are discussed in the next section.

On the other hand, the $b$ value in region B had been relatively large compared to the $b$ value in region A throughout the whole period before the $M$6.9 earthquake, although region B showed a rather large range in $b$ extending from 0.6 to 1.2 (Figure 5). It is also worth noting that the large-*b*-value state in region B has been continuing not only after the occurrence of the *M*6.9 earthquake on March 20, but also after the *M*6.8 earthquake on May 1. We suppose this result suggests that the differential stress in the eastern part of the source region of the 1978 off-Miyagi prefecture earthquake has been relatively low throughout the entire study period.

**Discussion**

By making maps and cross sections of the *b* value in the region around the foci of the March 20 and May 1, 2021, off-Miyagi prefecture earthquakes, we found that the *b* value in their focal areas had been considerably and consistently small and that the value became as low as 0.4-0.6 within a few years before their occurrence (Figures 2 and 3). The distinct small *b*-value spot corresponded to the eventual *M*6.9 hypocenter (Figures 2 and 3). This close match between the area of low and decreasing *b*-values and the eventual *M*6.9 and *M*6.8 hypocenters supports the idea that the *b* value may be a stress meter for the Earth's crust. Our finding also indicates that the stress before the *M*6.9 and *M*6.8 earthquakes had been high in the eventual focal areas, and the differential stress there had been heightened as time progressed. Here, it is to be noted that the coseismic slip of the 2011 Tohoku earthquake (orange contours in Figure 1a) stopped just short of the east side of the focus of the $M$6.9 earthquake and the afterslip of the 2011 Tohoku earthquake in 7 months (black contours in Figure 1a) had been relatively small in the focal areas of the two 2021 earthquakes, compared to the surrounding regions. Further, it is reported that the focus of the $M$6.9 earthquake was situated at the border point where the interplate slip during the period from 2012 through 2021 has been large on the northern side and notably smaller on the southern side (Tohoku University, 2021b).

The appearance of the notably small-$b$-value spot might have been related to the characteristics of the interplate slip events. We believe our results present a clear additional example to supplement previous retrospective studies that showed a correlation between patches of small $b$ values and sources of large earthquakes, e.g., at Izmit (Turkey), Parkfield and Ridgecrest (California), Tohoku and Kumamoto (Japan), and Iquique (Chile) (Wiemer and Wyss, 2002; Schorlemmer and Wiemer, 2005; Nanjo et al., 2012, 2016; Tormann et al., 2012, 2015; Schurr et al., 2014; Nanjo, 2020).

It is notable that the *b* value for the events after the *M*6.9 earthquake on March 20, which occurred mainly near the southern end of the slip area had been as small as 0.5-0.6 (blue) (bottom panel of Figure 2a and left panel of Figure 3b). We should have focused our attention more on the observation that the events after the *M*6.9 earthquake showed a small *b* value and the rupture of the quake had not covered the whole small-*b*-value area that had existed before the quake, although it is not certain if we could have foreseen the occurrence of the *M*6.8 earthquake before May 1.

The finding that the *b* value that had appeared to be small before the *M*6.9 earthquake was still low after the *M*6.8 earthquake is somewhat an enigma (Figure 5). This might mean that the two earthquakes had not fully unloaded the stress on the pre-existing asperity as interpreted for the Parkfield earthquake in 2004 (Tormann et al., 2012). However, we consider that such an interpretation cannot be applied to our case, because the rupture areas of the two earthquakes on March 20 and May 1 seem to cover almost entirely the small-*b*-value zone that had existed before their occurrence (bottom panel of Figure 2a). One possible explanation may be that some patches with high stress had remained unruptured, and events after the *M*6.8 earthquake have been occurring there. The results of seismic source analysis of the $M$6.9 and $M$6.8 earthquakes (Tohoku University, 2021a), which indicate a rather complex rupture process, seem to support this idea. Moreover, we would like to point out the occurrence of an $M$5.8 earthquake on April 18 at the far end of the rupture area of the earthquake on May 1 (Figure 5a and Figures S3a,c of Additional file 1). The sequential occurrence of the $M$6.9, $M$5.8 and $M$6.8 earthquakes in the small-$b$-value zone that had been observed before these earthquakes indicates that there had existed at least three high-stress asperities. This also seems to imply that additional smaller patches might have remained unruptured.

ERC (2021a), after a meeting held on March 22, commented that it was necessary to pay attention to the occurrence of another large earthquake that might result in a similar or even stronger seismic intensity during the period of one week, especially in a few days. We suppose that, in the background of this caution, ERC (2021a) concerned about the occurrence of the so-called off-Miyagi prefecture earthquake that has been occurring sequentially at intervals of about 38 years, and whose probability of occurrence within the next 30 years was estimated to be about 60-70% as of January 1, 2021 (ERC, 2021a,b). In addition, Nakata et al. (2016), based on numerical simulation, suggested that the time interval between the $M$~9 earthquake and the subsequent earthquake off the coast of Miyagi prefecture would become shorter than the average recurrence interval during the later stage of the $M$~9 earthquake cycle. As was pointed out in the Introduction, the two 2021 earthquakes occurred in the western part of the source region of the 1978 $M$7.4 off-Miyagi prefecture earthquake, and the focal area of the $M$6.8 earthquake is located west of the source region of the 2005 earthquake that occurred off the coast of Miyagi prefecture, fracturing the southern part of the source region of the 1978 earthquake (Figure 1b). Therefore, it might not be unreasonable for ERC (2021a) to have been anxious about the possibility of the occurrence of a large earthquake on the east side of the focal areas of the two 2021 earthquakes.

Concerning this anxiety, we would like to note that the $b$ value in the region had consistently been rather large before the two 2021 earthquakes and that the seismicity induced there by those earthquakes has been showing a large $b$ value as well (Figures 4 and 5). This indicates that stress in the region on the east side of the focal area of the two 2021 earthquakes had not been so high and the low-stress state has been continuing. Therefore, we conjecture that the probability of occurrence of a large earthquake in the adjacent region in the very near future may not be so large, although it is necessary to continue to watch for any signal that indicates change in local stress in the region.

**Conclusions**

This study revealed that the $b$ value in and around the focal areas of the $M$6.9 and $M$6.8 earthquakes that occurred off the Pacific coast of Miyagi prefecture, northeastern Japan, on March 20 and May 1, 2021, respectively, had been considerably low before their occurrence. The $b$ value in the vicinity of the $M$6.9 earthquake decreased to around 0.4 in the last few years. On the other hand, the $b$ value on the east side of the focal areas that corresponds to the eastern part of the source region of the 1978 off-Miyagi prefecture earthquake had been relatively large during the whole period that was investigated. This result implies that the stress in the region had not been as high as the stress in the focal areas of the two earthquakes in 2021 and that the low stress state there has been continuing.

# Declarations

### Ethics approval and consent to participate

Not applicable.

**Consent for publication**

Not applicable.

**List of abbreviations**

AO: Aomori

EMR: Entire-Magnitude-Range

ERC: Earthquake Research Committee

F-net: Full Range Seismograph Network of Japan

FU: Fukushima

GR law: Gutenberg-Richter's law

IB: Ibaraki

IW: Iwate

JMA: Japan Meteorological Agency

$M$: magnitude

$M_c$: Magnitude of completeness (or Completeness magnitude)

MI: Miyagi

NIED: National Research Institute for Earth Science and Disaster Resilience

**Availability of data and materials**

The JMA earthquake catalog was obtained from http://www.data.jma.go.jp/svd/eqev/data/bulletin/index_e.html. Coseismic slip of the 2021 $M$6.9 and $M$6.8 earthquakes determined by Keisuke Yoshida (Tohoku University, 2021a) could be obtained from https://www.static.jishin.go.jp/resource/monthly/2021/2021_04.pdf. Copyright permission for the $M$6.9 and $M$6.8 slip-models was obtained from Keisuke Yoshida. An updated version of these models was given in a submitting paper (Yoshida K, Matsuzawa T, Uchida N (2021) The 2021 Mw7.0 Miyagi-Oki earthquake, northeastern Japan, nucleated from deep plate boundary: Implications for the initiation of the M9 earthquake cycle. Submitted to Journal of Geophysical Research - Solid Earth). This preprint is available at https://doi.org/10.1002/essoar.10507585.1 (Accessed on August 31, 2021). Coseismic and postseismic slips of the 2011 Tohoku earthquake were obtained from Figure 12 of Ozawa et al. (2012). Coseismic slips of the 1978 $M$7.4 off-Miyagi prefecture earthquake and the 2005 $M$7.2 earthquake that occurred at off-Miyagi prefecture were obtained from Yamanaka and Kikuchi (2004) and Yaginuma et al. (2006), respectively. The upper surface of the Pacific plate (Nakajima and Hasegawa, 2006) was obtained from https://www.mri-jma.go.jp/Dep/sei/fhirose/plate/en.index.html. Focal mechanism catalog of F-net (Okada et al., 2004) was obtained from https://www.fnet.bosai.go.jp/top.php?LANG=en.

**Competing interests**

**Funding**

This study was partially supported by JSPS KAKENHI Grant Number JP 20K05050, the Tokio Marine Kagami Memorial Foundation, the Chubu Electric Power's research based on selected proposals, and the Ministry of Education, Culture, Sports, Science and Technology (MEXT) of Japan, under the Second Earthquake and Volcano Hazards Observation and Research Program (Earthquake and Volcano Hazard Reduction Research).

**Authors' contributions**

KZN and AY designed the study, KZN carried out analysis, and KZN and AY developed the manuscript. Both authors read and approved the final manuscript.

**Acknowledgements**

We would like to express our sincere thanks to the Editors (A. Kato and N. Uchida) and two anonymous reviewers for their very useful comments and suggestions that have improved the manuscript greatly. The authors thank the JMA for the earthquake catalog, T. Nagao for help with acquisition of this catalog, and K. Yoshida for copyright permission to use the $M$6.9 and $M$6.8 slip-models (Tohoku University, 2021a). The seismicity analysis software package ZMAP (Wiemer, 2001), used for Figures 2-5 and Figures S1-S6 of Additional file 1, was obtained from http://www.seismo.ethz.ch/en/research-and-teaching/products-software/software/ZMAP. The Generic Mapping Tools (Wessel et al., 2013), used for Figures 1-2, 4, and 5, and Figures S1 and S3-6 of Additional file 1, are an open-source collection (https://www.generic-mapping-tools.org).

**Authors' information**

K. Z. Nanjo

Present address: Global Center for Asian and Regional Research, University of Shizuoka, 3-6-1, Takajo, Aoi-ku, Shizuoka, 420-0839, Japan

Affiliations

Global Center for Asian and Regional Research, University of Shizuoka, 3-6-1, Takajo, Aoi-ku, Shizuoka, 420-0839, Japan

K. Z. Nanjo

Center for Integrated Research and Education of Natural Hazards, Shizuoka University, 836, Ohya, Suruga-ku, Shizuoka, 422-8529, Japan

K. Z. Nanjo & A. Yoshida

Institute of Statistical Mathematics, 10-3 Midori-cho, Tachikawa, Tokyo, 190-8562, Japan

K. Z. Nanjo

**Endnotes**

Not applicable.

**Additional information**

Additional file 1 includes Figures S1-S6. Additional files 2 and 3 include numerical data that list the longitude and latitude consisting of slip contours of the $M$6.9 and $M$6.8 earthquakes, respectively.

**References**

Aki K (1965) Maximum likelihood estimates of $b$ in the formula log$N$=$a$-$bM$ and its confidence limits. Bull Earthquake Res Inst, Univ Tokyo 43(2): 237-239. http://hdl.handle.net/2261/12198.

Earthquake Research Committee (2021a) Evaluation of the March 20, 2021, earthquake that occurred at off-Miyagi prefecture, released on March 22, 2021 (In Japanese). https://www.static.jishin.go.jp/resource/monthly/2021/20210320_miyagi_1.pdf. Accessed on August 31, 2021.

Earthquake Research Committee (2021b) Evaluation of the March 20, 2021, earthquake that occurred at off-Miyagi prefecture, released on April 9, 2021 (In Japanese).

https://www.static.jishin.go.jp/resource/monthly/2021/20210320_miyagi_2.pdf. Accessed on August 31, 2021.

Earthquake Research Committee (2021c) Evaluation of seismicity in April 2021 (In Japanese). https://www.static.jishin.go.jp/resource/monthly/2021/2021_04.pdf. Accessed on August 31, 2021.

Ghosh A, Newman AV, Thomas AM, Farmer GT (2008) Interface locking along the subduction megathrust from *b*-value mapping near Nicoya Peninsula, Costa Rica. Geophys Res Lett 35: L01301. doi:10.1029/2007GL031617.

Goebel THW, Schorlemmer D, Becker TW, Dresen G, Sammis CG (2013) Acoustic emissions document stress changes over many seismic cycles in stick-slip experiments. Geophys Res Lett 40: 2049-2054. doi:10.1002/grl.50507.

Gulia L, Wiemer S (2019) Real-time discrimination of earthquake foreshocks and

aftershocks. Nature 574, 193-199. doi:10.1038/s41586-019-1606-4.

Gutenberg B, Richter CF (1944) Frequency of earthquakes in California. Bull Seismological Soc Am 34(4), 185-188. doi:10.1785/BSSA0340040185.

Hirose, F., Nakamura A, Hasegawa A (2002) *b*-value variation associated with the rupture of asperities-Spatial and temporal distributions of *b*-value east off NE Japan (in Japanese). J Seismol Soc Jpn 55(3): 249–260. doi:10.4294/zisin1948.55.3_249.

Igarashi T, Matsuzawa T, Umino N, Hasegawa A. (2001) Spatial distribution of focal mechanisms for interplate and intraplate earthquakes associated with the subducting Pacific plate beneath the northeastern Japan arc: A triple-planed deep seismic zone. J Geophys Res 106(B2): 2177-2191. doi:10.1029/2000JB900386.

Japan Meteorological Agency (2012) The criterion of hypocenter determination process and the earthquake detection level after the 2011 off the Pacific coast of Tohoku

Earthquake (In Japanese). Report of the Coordinating Committee for Earthquake Prediction, Japan 87: 8-13. https://cais.gsi.go.jp/YOCHIREN/report/kaihou87/01_03.pdf. Accessed on August 31, 2021.

Lei X (2003) How do asperities fracture? An experimental study of unbroken asperities. Earth Planet Sci Lett 213(3-4): 347-359. doi:10.1016/S0012-821X(03)00328-5.

Mogi K (1963) The fracture of a semi-infinite body caused by an inner stress origin and its relation to the earthquake phenomena (Second Paper): The case of the materials having some heterogeneous structures. Bull Earthquake Res Inst, Univ Tokyo 41(3): 595-614. http://hdl.handle.net/2261/12128.

Nakajima J, Hasegawa A (2006) Anomalous low-velocity zone and linear alignment of seismicity along it in the subducted Pacific slab beneath Kanto, Japan: Reactivation of subducted fracture zone? Geophys Res Lett 33: L16309. doi:10.1029/2006GL026773.

Nakata R, Hori T, Hyodo M, Ariyoshi K (2016) Possible scenarios for occurrence of M ~ 7 interplate earthquakes prior to and following the 2011 Tohoku-Oki earthquake based on numerical simulation. Sci Rep 6: 25704. doi:10.1038/srep25704.

Nanjo KZ (2020) Were changes in stress state responsible for the 2019 Ridgecrest, California, earthquakes? Nat Commun 11: 3082. doi:10.1038/s41467-020-16867-5.

Nanjo KZ, Hirata N, Obara K, Kasahara K (2012) Decade-scale decrease in $b$ value prior to the $M$9-class 2011 Tohoku and 2004 Sumatra quakes. Geophys Res Lett 39: L20304. doi:10.1029/2012GL052997.

Nanjo KZ, Ishibe T, Tsuruoka H, Schorlemmer D, Ishigaki Y, Hirata N (2010) Analysis of the completeness magnitude and seismic network coverage of Japan. Bull Seismol Soc Am 100(6): 3261-3268. doi:10.1785/0120100077.

Nanjo KZ, Izutsu J, Orihara Y et al. (2016) Seismicity prior to the 2016 Kumamoto earthquakes. Earth Planets Space 68: 187. doi:10.1186/s40623-016-0558-2.

Nanjo KZ, Izutsu J, Orihara, Y, Kamogawa M, Nagao Y (2019) Changes in seismicity pattern due to the 2016 Kumamoto earthquakes identify a highly stressed area on the Hinagu fault zone. Geophys Res Lett 46(16): 9489-9496. doi:10.1029/2019GL083463.

Nanjo KZ, Yoshida A (2017) Anomalous decrease in relatively large shocks and increase in the $p$ and $b$ values preceding the April 16, 2016, $M$7.3 earthquake in Kumamoto, Japan. Earth Planets Space 69: 13. doi:10.1186/s40623-017-0598-2.

Nanjo KZ, Yoshida A (2018) A $b$ map implying the first eastern rupture of the Nankai Trough earthquakes. Nat Commun 9: 1117. doi:10.1038/s41467-018-03514-3.

Narteau C, Byrdina S, Shebalin P, Schorlemmer D (2009) Common dependence on stress for the two fundamental laws of statistical seismology. Nature 462: 642-645.

doi:10.1038/nature08553.

Okada Y, Kasahara K, Hori S, Obara K, Sekiguchi S, Fujiwara H, Yamamoto A (2004) Recent progress of seismic observation networks in Japan-Hi-net, F-net, K-NET and KiK-net. Earth Planets Space 56(8): xv-xxviii. doi:10.1186/BF03353076.

Ozawa S, Nishimura T, Munekane H, Suito H, Kobayashi T, Tobita M, Imakiire T (2012) Preceding, coseismic, and postseismic slips of the 2011 Tohoku earthquake, Japan. J Geophys Res 117(B7): B07404. doi:10.1029/2011JB009120.

Scholz CH (1968) The frequency-magnitude relation of microfracturing in rock and its relation to earthquakes. Bull Seismol Soc Am 58(1): 399-415.

Scholz CH (2015) On the stress dependence of the earthquake $b$ value. Geophys Res Lett 42(5): 1399-1402. doi:10.1002/2014GL062863.

Schorlemmer D, Hirata N, Ishigaki Y, Doi K, Nanjo KZ, Tsuruoka H, Beutin T, Euchner F (2018) Earthquake detection probabilities in Japan. Bull Seismological Soc Am 108(2): 702-717. doi:10.1785/0120170110.

Schorlemmer D, Wiemer S (2005) Microseismicity data forecast rupture area. Nature 434: 1086. doi:10.1038/4341086a.

Schorlemmer D, Wiemer S, Wyss M (2004) Earthquake statistics at Parkfield: 1. Stationarity of $b$ values. J Geophys Res 109: B12307. doi:10.1029/2004JB003234.

Schorlemmer D, Wiemer S, Wyss M (2005) Variations in earthquake-size distribution across different stress regimes. Nature 437: 539-542. doi:10.1038/nature04094.

Schurr B, Asch G, Hainzl S et al. (2014) Gradual unlocking of plate boundary controlled initiation of the 2014 Iquique earthquake. Nature 512: 299-302. doi:10.1038/nature13681.

Shi Y, Bolt BA (1982) The standard error of the magnitude-frequency *b* value. Bull Seismol Soc Am 72(5): 1677-1687. doi:10.1785/BSSA0720051677.

Smith WD (1981) The *b*-value as an earthquake precursor. Nature 289: 136-139. doi:10.1038/289136a0.

Sobiesiak M, Meyer U, Schmidt S, Götze H-J, Krawczyk CM (2007) Asperity generating upper crustal sources revealed by *b* value and isostatic residual anomaly grids in the area of Antofagasta, Chile. J Geophys Res 112: B12308, doi:10.1029/2006JB004796.

Suyehiro S, Asada T, Ohtake M (1964) Foreshocks and aftershocks accompanying a perceptible earthquake in central Japan - On the peculiar nature of foreshocks -. Papers Meteorol Geophys 15:71-88. doi:10.2467/mripapers1950.15.1_71.

Tohoku University (2021a) Coseismic slip distribution estimated from waveform inversion of the May 1, 2021, off Miyagi prefecture earthquake. In: ERC (ed) Evaluation of seismicity in April 2021, pp 18 (In Japanese). https://www.static.jishin.go.jp/resource/monthly/2021/2021_04.pdf. Accessed August 31, 2021.

Tohoku University (2021b) Space-temporal change in the slip rate. In: Report of the 231st Coordinating Committee for Earthquake Prediction, Japan, May 22-28, 2021 (In Japanese). https://cais.gsi.go.jp/YOCHIREN/activity/231/image231/030.pdf. Accessed August 31, 2021.

Tormann T, Enescu B, Woessner J, Wiemer S (2015) Randomness of megathrust earthquakes implied by rapid stress recovery after the Japan earthquake. Nat Geosci 8: 152-158. doi:10.1038/ngeo2343.

Tormann T, Wiemer S, Hardebeck JL (2012) Earthquake recurrence models fail when earthquakes fail to reset the stress field. Geophys Res Lett 39(18): L18310. doi:10.1029/2012GL052913.

Utsu T (1965) A method for determining the value of $b$ in a formula log$n$=$a$-$bM$ showing the magnitude-frequency relation for earthquakes. Geophys Bull Hokkaido Univ 13: 99-103 (in Japanese). doi:10.14943/gbhu.13.99.

Utsu T (1992) On seismicity. In: Report of the Joint Research Institute for Statistical Mathematics, Inst Stat Math Tokyo. vol 34, pp 139-157.

Utsu T (1999) Representation and analysis of the earthquake size distribution: A historical review and some approaches. Pure Appl Geophys 155(2): 509–535. doi:10.1007/s000240050276.

Wessel P, Smith WHF, Scharroo R, Luis J, Wobbe F (2013) Generic Mapping Tools: Improved version released. EOS Trans AGU 94: 409-410. doi:10.1002/2013EO450001.
35

Wiemer S (2001) A software package to analyze seismicity: ZMAP. Seismological Res Lett 72(3): 373-382. doi:10.1785/gssrl.72.3.373.

Wiemer S, Wyss M (2002) Mapping spatial variability of the frequency-magnitude distribution of earthquakes. Adv Geophys 45: 259-302. doi:10.1016/S0065-2687(02)80007-3.

Woessner J, Wiemer S (2005) Assessing the quality of earthquake catalogues: Estimating the magnitude of completeness and its uncertainty. Bull Seismological Soc Am 95(2): 684-698. doi:10.1785/0120040007.

Yabe Y (2003) Frictional property of plate interface east off NE Japan inferred from spatial variation in *b*-value. Bull Earthquake Res Inst, Univ Tokyo 78(1): 107-111. http://hdl.handle.net/2261/5737.

Yaginuma T, Okada T, Yagi Y, Matsuzawa T, Umino N, Hasegawa A (2006) Coseismic slip distribution of the 2005 off Miyagi earthquake (M7.2) estimated by inversion of teleseismic and regional seismograms. Earth Planets Space 58: 1549-1554. doi:10.1186/BF03352659.

Yamanaka Y, Kikuchi M (2004) Asperity map along the subduction zone in northeastern Japan inferred from regional seismic data. J Geophys Res 109(B7): B07307. https://doi.org/10.1029/2003JB002683.

**Figure legends**

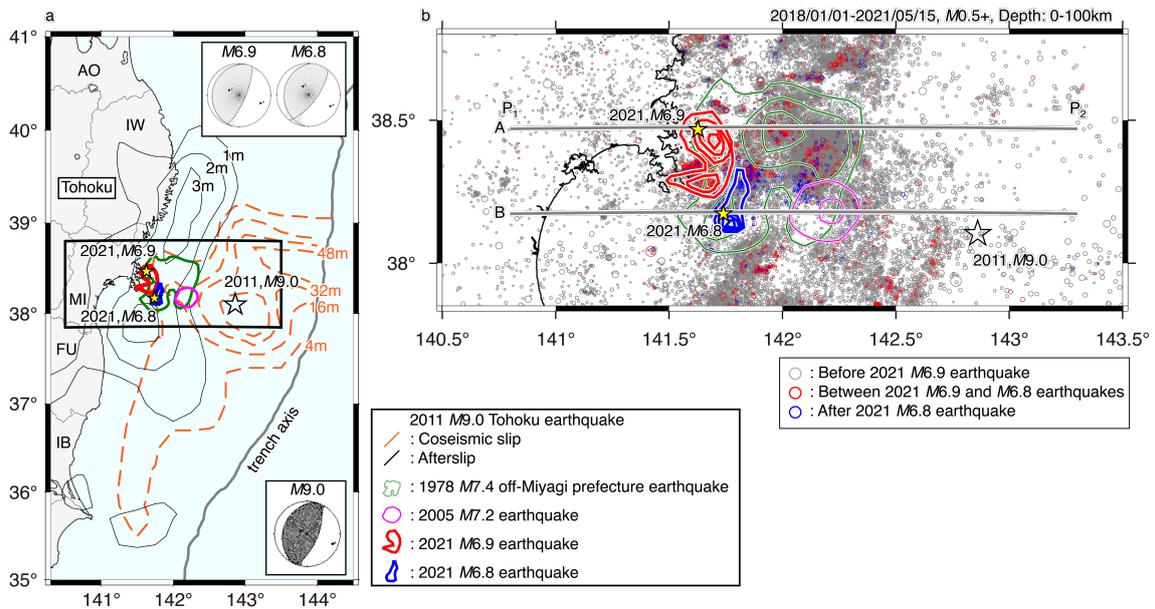

Figure 1. Earthquakes off the Pacific coast of Tohoku district. (a) Yellow stars represent the epicenters of the 2021 *M*6.9 and *M*6.8 earthquakes; open star shows the epicenter of the 2011 *M*9.0 Tohoku earthquake. Focal mechanism for these earthquakes, obtained by using the focal mechanism catalog of F-net (Full Range Seismograph Network of Japan) by NIED: National Research Institute for Earth Science and Disaster Resilience (Okada et al., 2004), is displayed as a beach ball symbol. Red and blue regions represent preliminary-determined slip areas of the 2021 *M*6.9 and *M*6.8 earthquakes, respectively (Tohoku University, 2021a). The coseismic slip distribution of the 2011 Tohoku

earthquake (Ozawa et al., 2012) is shown by orange contours with an interval of 16 m, except for the interval of 12 m between 4-meter contours and 16-meter contours. Black contours with an interval of 1 m indicate the afterslip distribution of the same earthquake (Ozawa et al., 2012). Green and purple regions show coseismic slip areas of the 1978 *M*7.4 off-Miyagi prefecture earthquake (Yamanaka and Kikuchi, 2004) and the 2005 *M*7.2 earthquake that occurred at off-Miyagi prefecture (Yaginuma et al., 2006), respectively. Black rectangle shows the study region in (b). AO (Aomori), IW (Iwate), MI (Miyagi), FU (Fukushima), IB (Ibaraki) are abbreviations of prefecture names. (b) Seismicity (depth 0-100 km, $M \geq 0.5$) in and after 2018. Cross sectional views along the line segments from $P_1$ to $P_2$, indicated by "A" and "B", are shown in Figure 3. Red and blue contours with an interval of 0.4 m indicate the coseismic slip distributions of the *M*6.9 and *M*6.8 earthquakes, respectively (Tohoku University, 2021a). We traced each slip contour from the original (Tohoku University, 2021a), created numerical data that list the longitude and latitude consisting of the slip contours (Additional files 2 and 3 for the *M*6.9 and *M*6.8 earthquakes, respectively), and drew the contours using the numerical data.

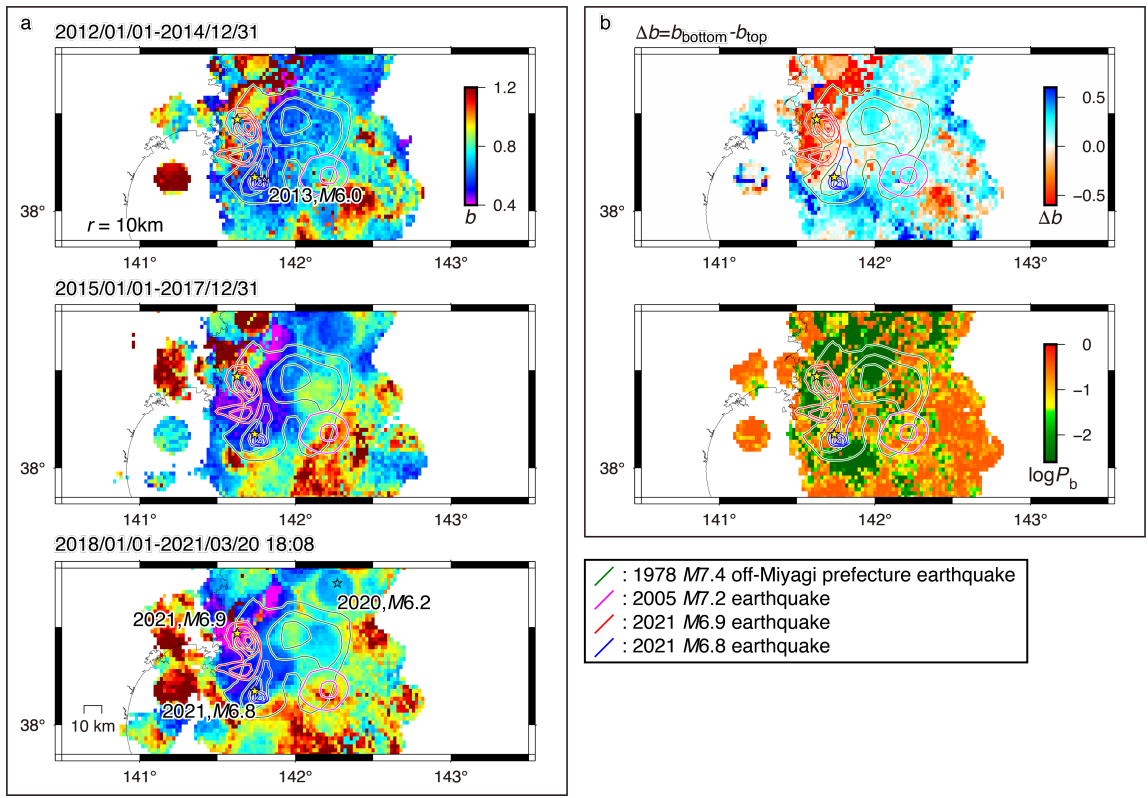

Figure 2. *b*-value analysis based on seismicity before the 2021 *M*6.9 earthquake. (a) Maps of *b* values obtained from seismicity during the three periods; top panel: from 2012 to 2014, middle panel: from 2015 to 2017, and bottom panel: from 2018 to immediately before the 2021 *M*6.9 earthquake. In making these maps, we selected earthquakes along the plate interface (Nakajima and Hasegawa, 2006): earthquakes were chosen if their depths were within a range from 5 km above the interface to 15 km below it. We calculated the *b* value and simultaneously $M_c$ (Figures S1a-c of Additional file 1) for each grid node (0.02° spacing) selecting all events within a search radius of 10 km. Other symbols are the same as in Figure 1. Epicenters of $M \geq 6.0$ earthquakes that occurred in the corresponding periods are shown by open stars. (b) Top panel: $\Delta b$ (= $b_{bottom}$ - $b_{top}$), the difference in $b$ values between the periods 2012-2014 ($b_{top}$: top panel in a) and 2018-2021 ($b_{bottom}$: bottom panel in a). Bottom panel: $\log P_b$, the logarithm of the probability that $b_{bottom}$ is not different from $b_{top}$.

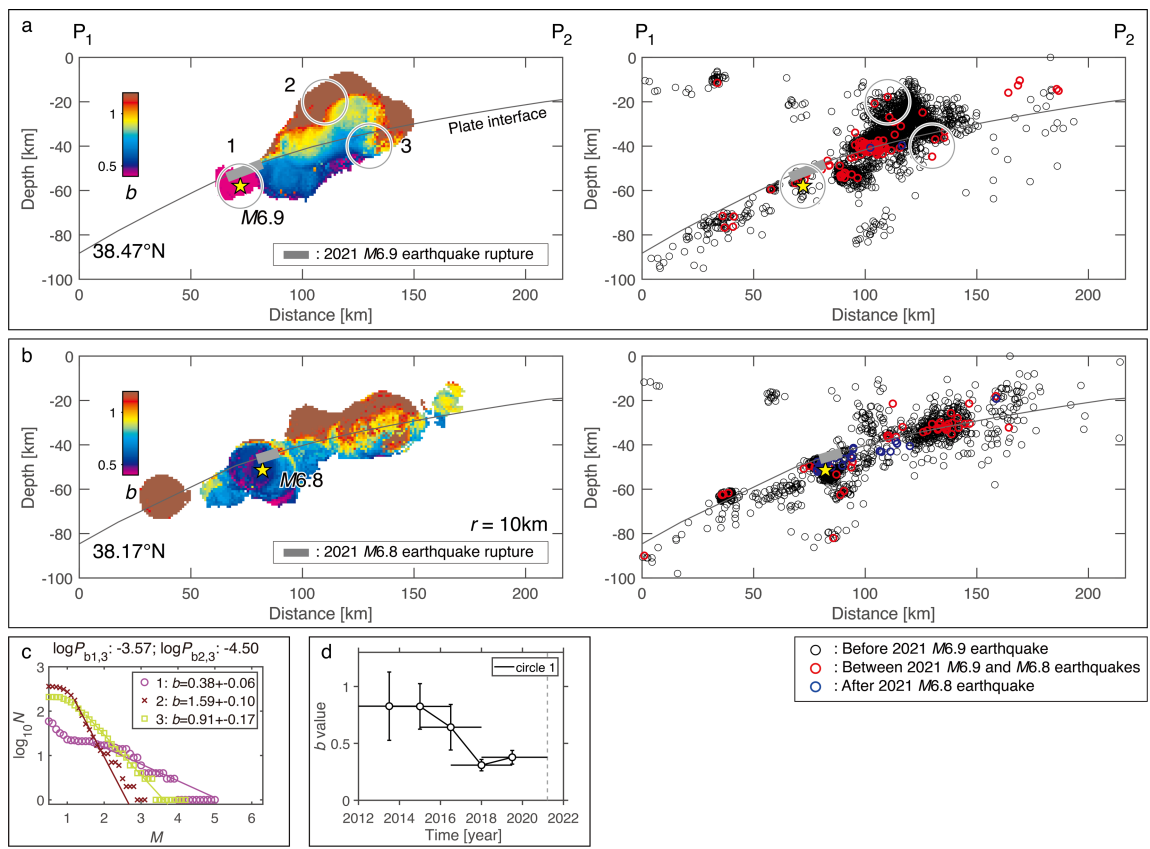

Figure 3. Cross sectional views of $b$ values and seismicity since 2018 until the 2021 $M$6.9 earthquake. (a) Left panel: $b$ values computed based on seismicity along the line

segment "A" in Figure 1b. To create this figure, we project sampled earthquakes onto a vertical plane extending from $P_1$ to $P_2$ in Figure 1b, roughly perpendicular to the Japan trench axis (Figure 1a). At each grid node (1 km grid spacing), we select all events within a search radius of 10 km and considered a swath width of 10 km, resulting, at each grid node, in a cylindrical sample volume of 20 km diameter and 10 km depth. Black curve indicates the upper surface of the Pacific plate (Nakajima and Hasegawa, 2006). Grey segments indicate the rupture areas of the 2021 *M*6.9 and *M*6.8 earthquakes (Tohoku University, 2021a) (hypocenters indicated by stars). Right panel: black, red, and blue circles represent seismicity before the 2021 *M*6.9 earthquake, between the 2021 *M*6.9 and *M*6.8 earthquakes, and after the 2021 *M*6.8 earthquake, respectively. (b) Same as (a) for the cross-section along the line segment "B" in Figure 1b. (c) Events within the three circles of radius $r = 10$ km, indicated by "1", "2", and "3", are used to show a fit of the GR law to observations. Color corresponds to the *b* value indicated by the color bar. The uncertainty estimates are according to Shi and Bolt (1982). $\log P_{b1,3}$ is the logarithm of the probability that the *b* values for the circles 1 and 3 are not different and $\log P_{b2,3}$ is the same as that for circles 2 and 3. (d) Plot of *b* values as a function of time for circle 1.

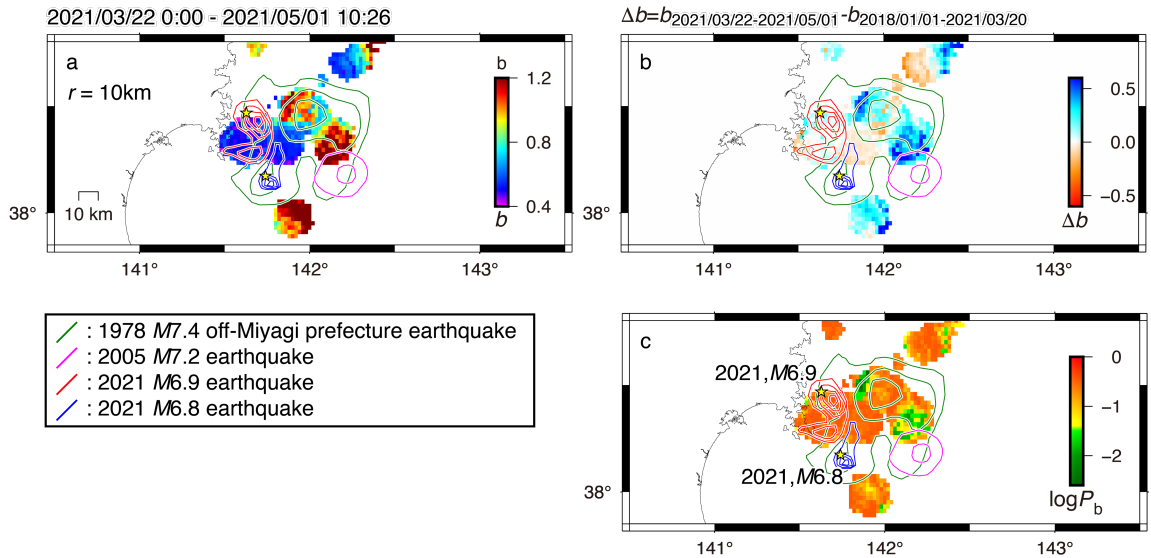

Figure 4. *b*-value analysis for the period between the *M*6.9 and *M*6.8 earthquakes. (a) Map of *b* values obtained from the seismicity for the period from March 22, at 0:00 to May 1, at 10:26 ($b_{2021/03/22-2021/05/01}$), where we do not consider the period (March 20, at 18:10 - March 21, 23:59) from immediately after the *M*6.9 March 20 earthquake to the end of March 21, avoiding the time period with large temporal variation in *b*. Other symbols are the same as in Figure 1. (b) Δ*b* (= $b_{2021/03/22-2021/05/01}$ - $b_{2018/01/01-2021/03/20}$), subtraction of the *b* value in the bottom panel in Figure 2a ($b_{2018/01/01-2021/03/20}$), from the *b* value in (a) ($b_{2021/03/22-2021/05/01}$). (c) log$P_b$, the logarithm of the probability that $b_{2018/01/01-2021/03/20}$ and $b_{2021/03/22-2021/05/01}$ are not different.

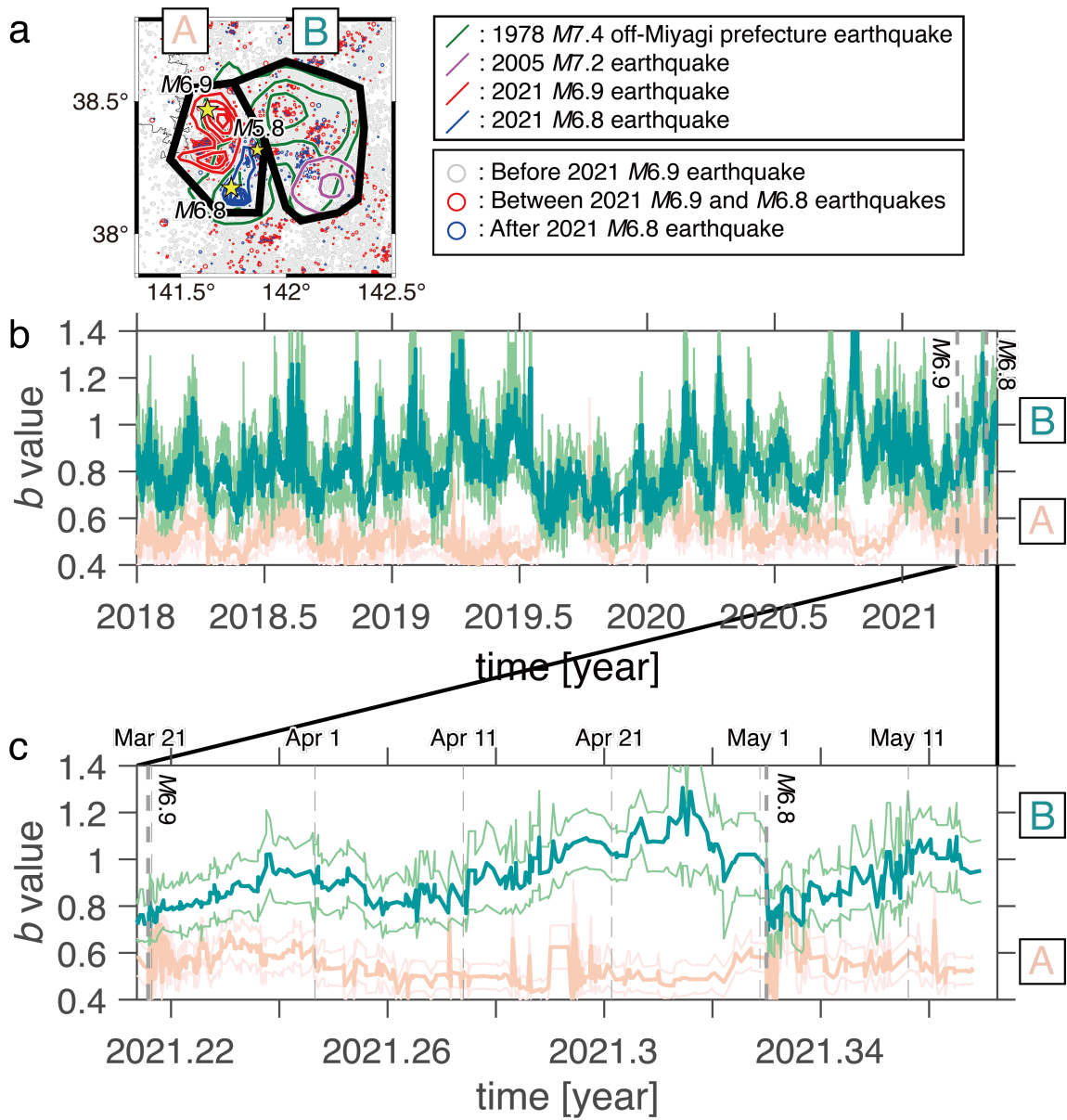

Figure 5. Time series of $b$ values. (a) Map showing the regions "A" and "B" considered in (b,c). Stars indicates the epicenters of the $M$6.9 March 20 earthquake, the $M$5.8 April 18 earthquake, and the $M$6.8 May 1 earthquake. Other symbols are the same as in Figure 1. (b) Plot of $b$ values as a function of time, as obtained from seismicity data

along the plate boundary for the period since 2018. We used a moving window approach, whereby the window covered 100 events. The uncertainty estimates are according to Shi and Bolt (1982). Thick vertical dashed lines indicate moment of the $M$6.9 and $M$6.8 earthquakes. (c) Zoom-in figure for the period from immediately before the 2021 $M$6.9 earthquake to May 15, 2021. Thin vertical dashed lines show March 21, April 1, 11, and 21, and May 1 and 11.